\documentclass[preprint]{revtex4}
\usepackage{epsfig}
\usepackage{latexsym,amsmath,amssymb,amstext}
\usepackage{float}
\usepackage{graphicx}
\usepackage{mathrsfs}

\newcommand\bef{\begin{figure}}
\newcommand\eef[1]{\label{fg:#1}\end{figure}}
\newcommand\beq{\begin{equation}}
\newcommand\eeq[1]{\label{#1}\end{equation}}
\newcommand\beqa{\begin{eqnarray}}
\newcommand\eeqa[1]{\label{#1}\end{eqnarray}}
\newcommand\bet{\begin{table}}
\newcommand\eet[1]{\label{tb:#1}\end{table}}

\newcommand\fgn[1]{Figure \ref{fg:#1}}
\newcommand\eqn[1]{Eq.\ (\ref{#1})}
\newcommand\scn[1]{Section \ref{sec:#1}}

\newcommand\ie{{\sl i.e.\/}}

\newcommand{\mn}{{\rm min}}
\newcommand{\mx}{{\rm max}}
\newcommand{\sun}{SU($N$)\ }
\newcommand{\cas}{C^{\scriptscriptstyle{(2)}}_r}

\begin{document}
\title{The equation of state of two dimensional Yang-Mills theory}
\author{Nikhil\ \surname{Karthik}}
\email{nkarthik@fiu.edu}
\affiliation{Department of Physics, Florida International University, Miami, FL 33199.}
\author{Rajamani\ \surname{Narayanan}}
\email{rajamani.narayanan@fiu.edu}
\affiliation{Department of Physics, Florida International University, Miami, FL 33199.}

\begin{abstract}
We study the pressure, $P$, of \sun gauge theory on a two-dimensional
torus as a function of area, $A=l/t$.  We find a  cross-over scale
that separates the system on a large circle from a system on a small
circle at any finite temperature. The cross-over  scale approaches
zero with increasing $N$ and the cross-over becomes a first order
transition as $N\to\infty$ and $l\to 0$ with the limiting value of
$\frac{2Pl}{(N-1)t}$ depending on the fixed value of $Nl$.
\end{abstract}

\maketitle

\section{Introduction}
\label{sec:intro}

The partition function for \sun gauge theory on a 2d torus with spatial extent $l$
and temperature $t$ is only a function of the area, $A=l/t$, and is given by
\cite{Migdal:1975zg}
\beq
Z_N(A)=\sum_r \exp\left(-\frac{\cas l}{Nt}\right),
\eeq{hkz}
where  $\cas$ is the value of Casimir in the representation $r$.
One can arrive at (\ref{hkz}) by taking the continuum limit of a lattice
formalism on a finite lattice~\cite {Kiskis:2014lwa}. 
The asymptotic behavior at large $N$ was studied in~\cite{Gross:1993hu}
where only representations with $\cas$ of $\mathscr{O}(N)$ dominate.
Since the partition
function is a sum over string like states with energies proportional to
the spatial extent, $l$, the pressure given by
\beq
P\equiv t\frac{\partial}{\partial l}\ln Z = \frac{\partial}{\partial A}\ln Z = -\frac{1}{N}\langle\cas\rangle,
\eeq{pressure}
is negative.

The partition function for SU(2) is simple and given by
\beq
Z = \sum_{\lambda=0}^\infty e^{-\frac{\left(\lambda^2 +2 \lambda\right)A}{4}} = \frac{1}{2} e^{\frac{A}{4}} \left [
\sum_{\lambda=-\infty}^\infty e^{-\frac{\lambda^2A}{4}} -1\right] =
\frac{1}{2} e^{\frac{A}{4}} \left[ \sqrt{\frac{4\pi}{A}} \sum_{\lambda=-\infty}^\infty e^{-\frac{4\pi^2\lambda^2}{A} }-1\right].
\eeq{partsu2}
The asymptotic behavior of the equation of state is
\beq
\frac{Pl}{t} = -\frac{3}{4} \frac{l}{t} e^{-\frac{3l}{4t}}\quad\text{as}\quad  l\to\infty,
\eeq{largea}
and
\beq
\frac{Pl}{t} = -\frac{1}{2} \quad\text{as}\quad l\to 0.
\eeq{smalla}
The behavior at large $l$ is dominated by a few low lying energy states
where as the behavior at small $l$ comes from a sum over all states and
 could be interpreted as the equipartition limit with
the number of degrees of freedom being 1 for SU(2).  The cross-over
from the behavior on a large circle to a small circle is shown in \fgn{su2}.
\bef
\begin{center}
\includegraphics[scale=0.8]{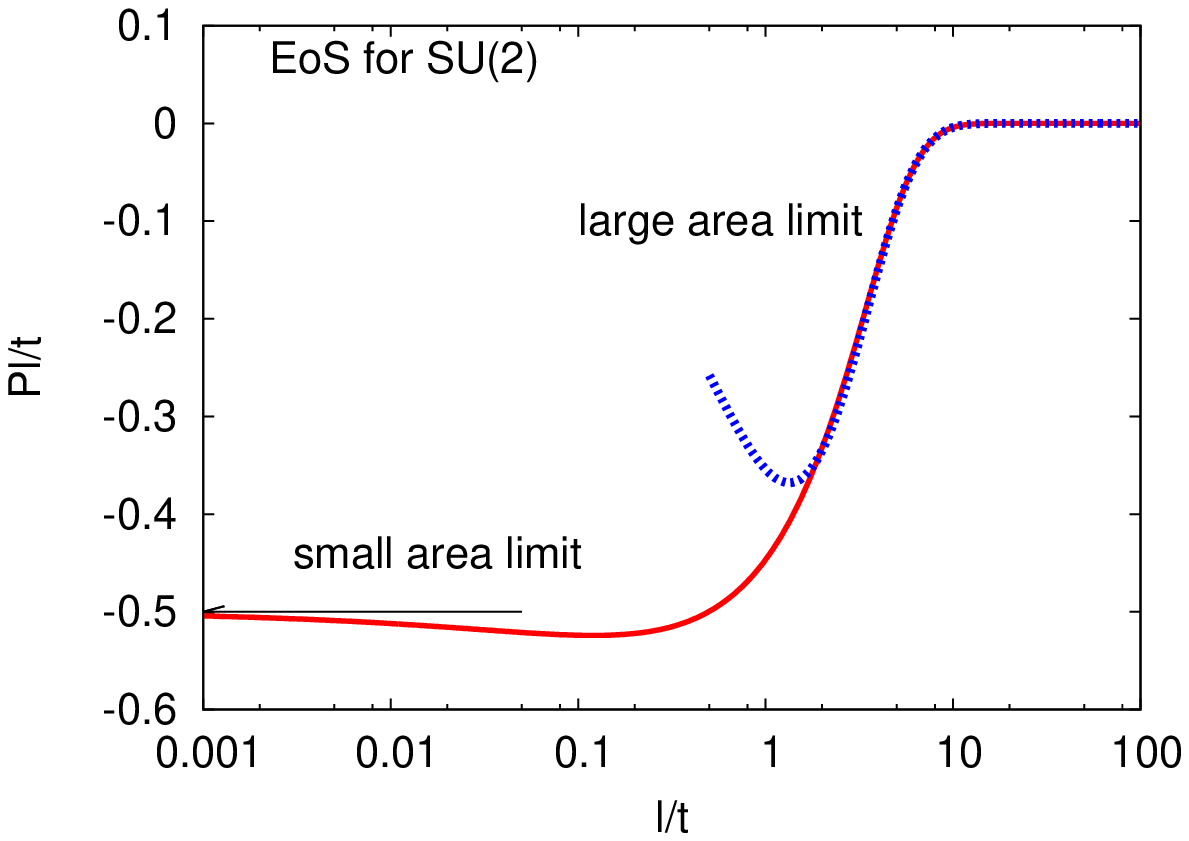}
\end{center}
\caption{The equation of state for SU(2) gauge theory on a two-dimensional
torus is shown as the solid curve. The asymptotic values of
$Pl/t$ at small area is -0.5. At very large area, $Pl/t$ behaves as
$0.75\exp(-0.75l/t)l/t$, which is shown as the dotted curve.
There is a cross-over between the two limits.}
\eef{su2}

Expecting that the equipartition limit is given by
\beq
\frac{Pl}{t}=-\frac{N-1}{2} \quad\text{as}\quad l\to 0,
\eeq{smallaN}
for all $N$, we define
\beq
Q(\alpha)\equiv -\frac{2Pl}{(N-1)t}; \quad\text{with}\quad \alpha=\frac{Nl}{t},
\eeq{Q}
and study this quantity in this paper.

\section{Summary of results}

We will show the following results in this paper using a numerical
simulation of the partition function in \eqn{hkz}:
\begin{enumerate}
\item $Q(\alpha)$ falls on a universal curve as $N\to\infty$.
\item $Q(\alpha)$ goes to zero as $\alpha$ goes to infinity.  This result
implies that the pressure at infinite $N$ is zero for all $l$ at any
$t$ as long as one takes $N\to\infty$ keeping $l$ and $t$ finite and
is consistent with physics being independent of temperature and spatial
extent in the infinite $N$ limit~\cite{Gross:1980he,Eguchi:1982nm}.
\item $Q(\alpha)$ goes to unity as $\alpha$ goes to zero. This limit is
reached from a finite $l$ and $t$ only at finite $N$.
\item There is a cross-over point defined as a peak in the susceptibility,
\beq
\chi=A\frac{\partial}{\partial A}Q=\alpha\frac{\partial}{\partial\alpha}Q.
\eeq{sus}
\begin{enumerate}
\item The large $l$ side of the cross-over is dominated by representations
where $\cas$ are of $\mathscr{O}(N)$. This is the case of interest for
all non-zero $l$ at infinite $N$ and studied in~\cite{Gross:1993hu}.
\item The small $l$ side of the cross-over is dominated by representations
where $\cas$ are of $\mathscr{O}(N^2)$.
\end{enumerate}
\item Since the value of $Q$ at infinite $N$ and $l=0$ (or equivalently
$t=\infty$) depends on the approach to the limit, $N\to\infty$ and
$l\to 0$, there is a first order transition confirming the argument
in~\cite{McLerran:1985uh}.
\end{enumerate}

\section{Properties of Casimir for SU(N)}
\label{sec:casimir}
The representations of \sun are specified by the sequence of integers
$\Lambda_r=\left(\lambda_1,\lambda_2,\ldots,\lambda_{N-1}\right)$,
subjected to the ordering $\lambda_i \ge \lambda_{i+1}$ and the value
of $\cas$ is 
\beq
\cas=\sum_{i=1}^{N-1}\lambda_i^2-\sum_{i=1}^{N-1}i\lambda_i-\frac{\lambda^2}{N}+(N+1)\lambda\quad\text{where}\quad \lambda=\sum_{i=1}^{N-1}\lambda_i.
\eeq{casimir} 
The maximum and the minimum value of Casimir, given the constraint
that $\lambda$ has to be kept fixed, would be used in the subsequent
sections. The representation with the maximum value of $\cas$ for a
given $\lambda$ is given by
\beq
\Lambda_\mx=(\lambda,0,\ldots,0).
\eeq{maxseq}
The minimum value of $\cas$ is given by the sequence $\Lambda_\mn$:
\beq
\lambda_i =
\left\{
	\begin{array}{ll}
		\lfloor\frac{\lambda}{N-1}\rfloor+1  & \mbox{if } i\le k\equiv\lambda-(N-1)\lfloor\frac{\lambda}{N-1}\rfloor \\
		\lfloor\frac{\lambda}{N-1}\rfloor & \mbox{if } i>k.
	\end{array}
\right.
\eeq{minseq}
To prove that the two sequences extremize the Casimir,
note that the Casimir decreases under the transformation
$(\lambda_1,\lambda_2,\ldots,\lambda_i,\ldots,\lambda_j,\ldots,\lambda_{N-1})$ to
$(\lambda_1,\lambda_2,\ldots,\lambda_i-1,\ldots,\lambda_j+1,\ldots,\lambda_{N-1})$
for $j>i$, provided this transformation is allowed. Such a transformation
is not possible for $\Lambda_\mn$.  Similarly, the reverse of that
transformation is not possible on $\Lambda_\mx$. One can prove by
contradiction that $\Lambda_\mn$ and $\Lambda_\mx$ are unique to satisfy
these properties.
\bef
\begin{center}
\includegraphics[scale=0.8]{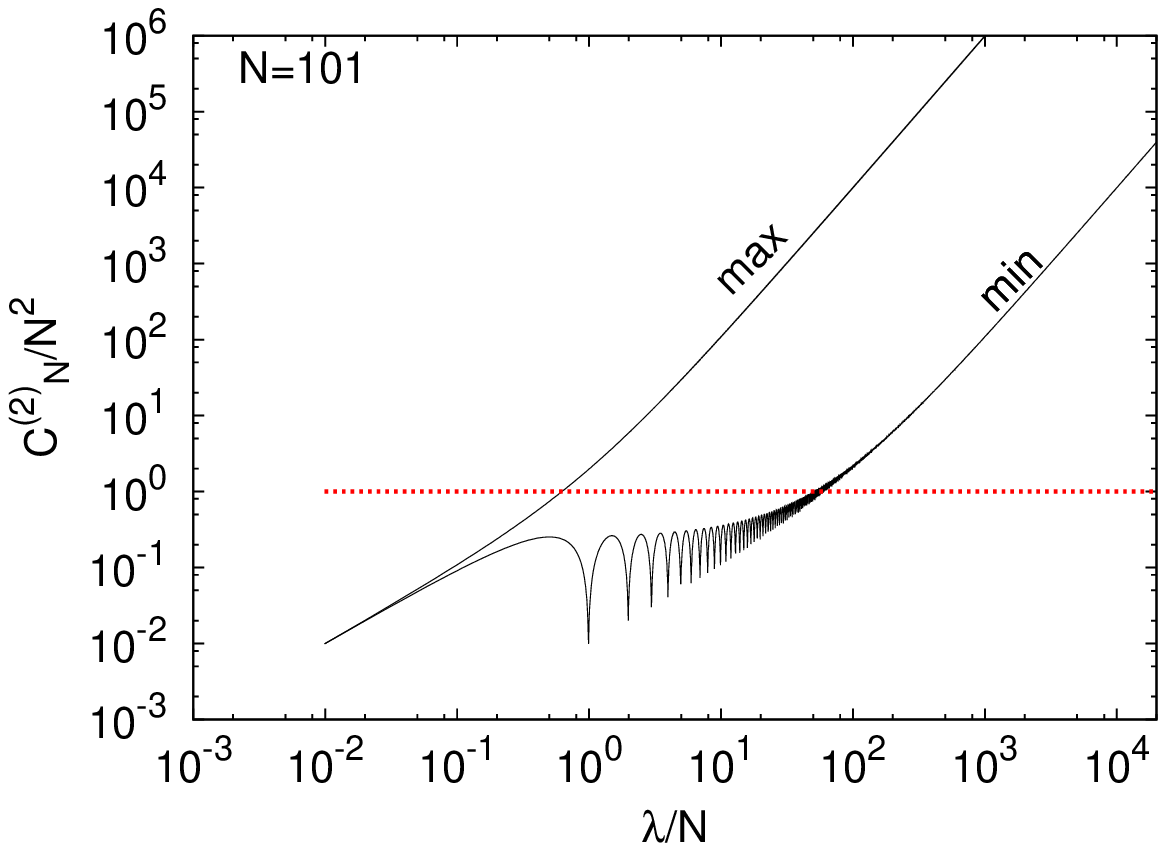}
\end{center}
\caption{Behaviour of Casimir as a function of $\lambda$. The upper 
solid curve is the maximum value of Casimir given a value of $\lambda$,
as a function of $\lambda$. Similarly, the lower solid  curve is
the minimum value of Casimir given a value of $\lambda$, as a function
of $\lambda$. The dotted line is where $\cas=N^2$.}
\eef{minmax}

We have shown the behaviour of the maximum and the minimum value of
$\cas$ as a function of $\lambda$ in \fgn{minmax}.  The minimum of $\cas$
shows a quasi-periodic behaviour, with troughs at $\lambda=q N$ for integer
$q$. The values of Casimir at these troughs are
\beq
C_\mn=N\left(1+\left\lfloor \frac{q}{N-1} \right\rfloor \right) \left( 2q - \left\lfloor \frac{q}{N-1} \right\rfloor (N-1) \right),
\eeq{minmin}
whose dependence on $N$ is linear for $q$ between two multiples of $(N-1)$
 and is quadratic for $q$ that are multiples of $(N-1)$.
On very large circles (or at very low temperatures), one would expect that only the
excitations around these troughs at small $q$ would be important. On very small circles (or at
very high temperatures), large values of $q$ would become accessible,
where all possible Casimir are $\mathscr{O}(N^2)$. This is the region
above the red dotted line in \fgn{minmax} where $\cas$ is larger than
$N^2$. Qualitatively, this is the difference one might expect between
the low and high temperature phases.

\section{Heat-bath algorithm}
\label{sec:method}

We simulated the partition function in \eqn{hkz} by updating $\Lambda_r$
by the heat-bath algorithm.  Each heat-bath update is a sequence of local
updates from $\lambda_1$ to $\lambda_{N-1}$, in that order, such that
the ordering of $\lambda_i$ is preserved.  For the local update of
$\lambda_i$, the probability distribution of $\lambda_i$
is given by a discrete version of the Gaussian distribution
\beq
T\left(\lambda_i\right)\propto e^{-(\lambda_i-\mu_i)^2/2\sigma^2},
\eeq{transition}
subject to the condition $\lambda_{i+1}\le\lambda_i\le\lambda_{i-1}$
for $i>1$ and $\lambda_{2} \le \lambda_1$.  The $\mu_i$ and $\sigma_i$
for the above discrete Gaussian distribution are functions of the rest
of the $\lambda_i$'s forming the heat-bath:
\beq
\mu_i=\frac{\overline{\lambda}+N\left(\frac{2i-N-1}{2}\right)}{N-1}\quad\text{and}\quad \sigma^2=\frac{N^2}{2A\left(N-1\right)},
\eeq{distribution}
where $\overline{\lambda}=\sum_{j\ne i}\lambda_j$.  For $i>1$, the set of
allowed values for $\lambda_i$ is bounded from above and below. Hence,
we included all the allowed possibilities weighted by \eqn{transition}
as candidates for the update. Since  \eqn{distribution}, along with the
inequality $\lambda-\lambda_1<(N-2)\lambda_2$, implies that $\mu_1 <
\lambda_2$, the probability for $\lambda_1$ is a monotonically decreasing
function. This enables one to put an upper cut-off on $\lambda_1$.
In our calculation, we used an upper cut-off of $\lambda_2+3\sigma$. We
also checked that changing this value to $\lambda_2+10\sigma$ does not
cause any statistically significant changes.  Since a representation
$r$ and its conjugate representation $\overline{r}$ have the same
Casimir, one can do an over-relaxation step by a global update
$\lambda^\prime_i=\lambda_1-\lambda_{N-i+1}$.

In our simulations, the successive measurements were separated by 100
iterations of 2 heat-bath and 1 over-relaxation steps. The first 2000
measurements were discarded for thermalization. In this way, we collected
$10^4$ configurations of $\Lambda_r$ at all area and $N$.

\bef
\begin{center}
\includegraphics[scale=0.8]{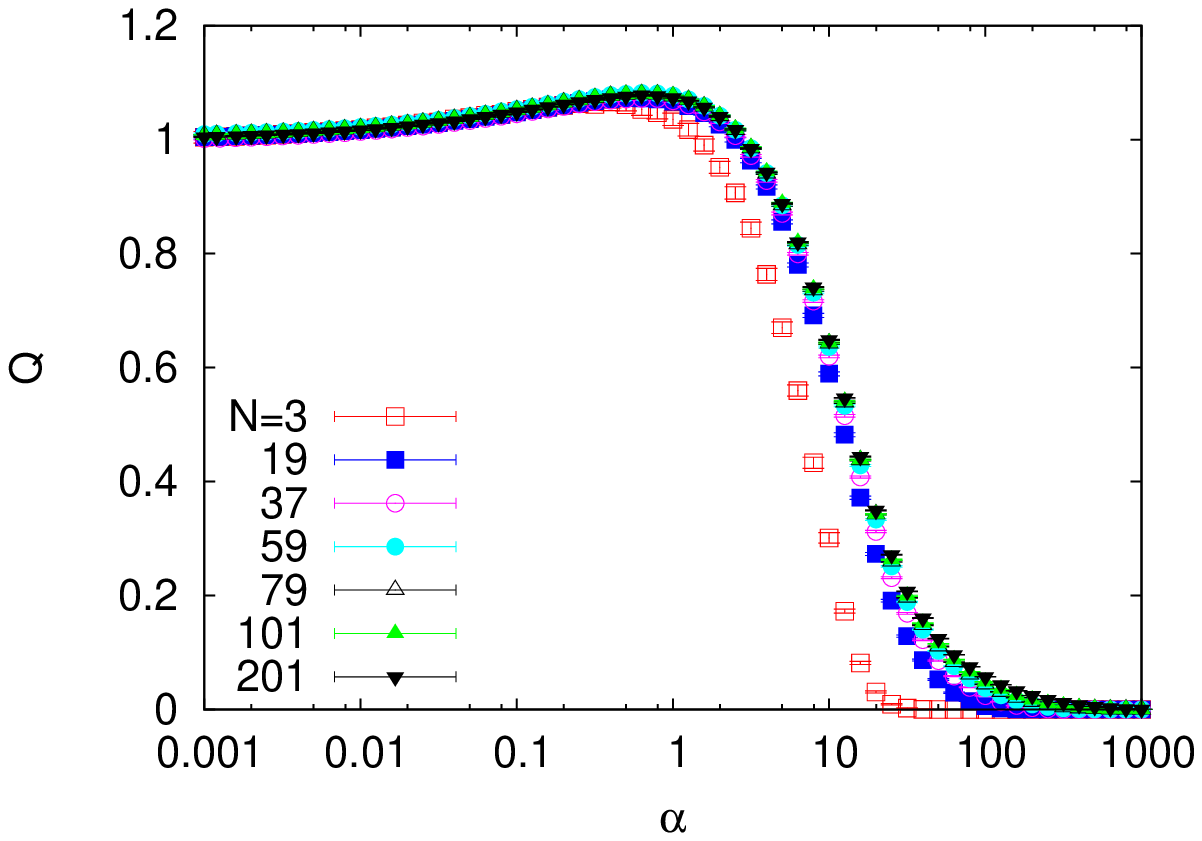}
\includegraphics[scale=0.8]{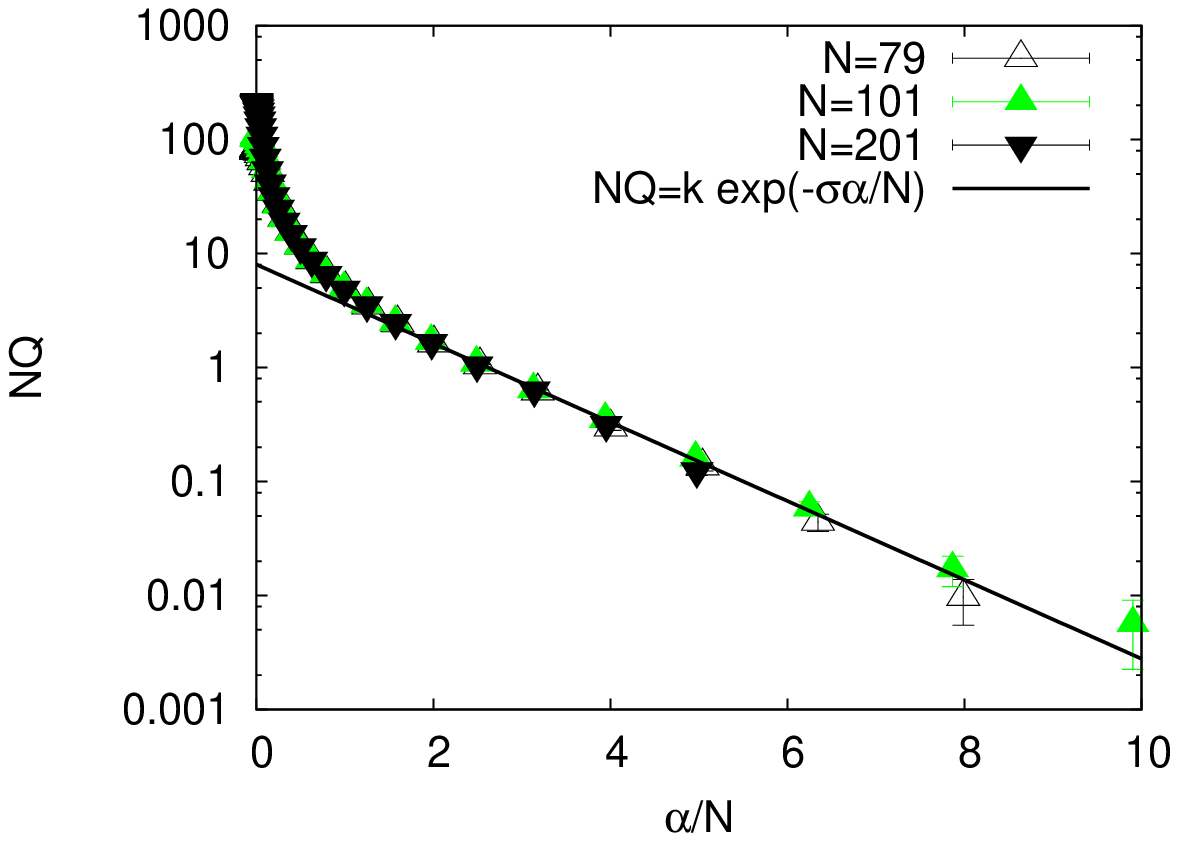}
\end{center}
\caption{$Q$ as a function of the scaled area $\alpha=NA$ is shown in the
top panel. It is seen that $Q$ as a function of $\alpha$ has a large-$N$
limit. For very small values of $\alpha$, $Q$ approaches $1$. In the
bottom panel, the large area behaviour of $Q$ in the large-$N$ limit is
shown. In this case, $QN$ behaves as $\exp\left(-\sigma\alpha/N\right)$.}
\eef{press}

\section{Results}
\label{sec:results}
In the top panel of \fgn{press}, we show the behaviour of $Q$ as a
function of the scaled area $\alpha$ for various values of $N$. The
important thing to notice is that $Q$ has a large-$N$ limit when
plotted as a function of $\alpha$. For $\alpha\ll 1$, $Q$ seems to
approach $1$ for all $N$.  This is in agreement with our intuition
based on the equipartition theorem. The non-trivial observation is
that this cross-over to the equipartition limit happens at a finite
value of $\alpha$ in the large-$N$ limit. For $\alpha\gg 1$, $Q$ seems
to behave as $N^{-1}\exp\left(-\sigma \alpha/N\right)$ for a constant
$\sigma\approx 0.81$ in the large-$N$ limit. This is shown in the bottom
panel of \fgn{press}.  Thus, it can also be seen as a cross-over from
strong-coupling regime, which has a scale $\sigma$, to the weak-coupling
regime with no underlying scale.

We determined the cross-over point $\alpha_c$ using the peak-position
of the susceptibility $\chi$, after interpolating using multi-histogram
reweighting.  We show $\chi$ as a function of $\alpha$ in \fgn{susfig}
for various $N$.  The susceptibility also has a large-$N$ limit when
plotted as a function of $\alpha$. The peak positions of susceptibility for
$N>19$ agree within errors, giving us an estimate $\alpha_c=12.1(2)$.
This implies that the cross-over area $A_c=\alpha_c/N$ shifts to smaller
values at larger $N$.  The width of the susceptibility when expressed in
terms of the area $A$ decreases inversely as $N$. This is characteristic
of finite volume scaling near a first order phase transition, with the
large-$N$ limit replacing the thermodynamic limit in this case.
\bef
\begin{center}
\includegraphics[scale=0.8]{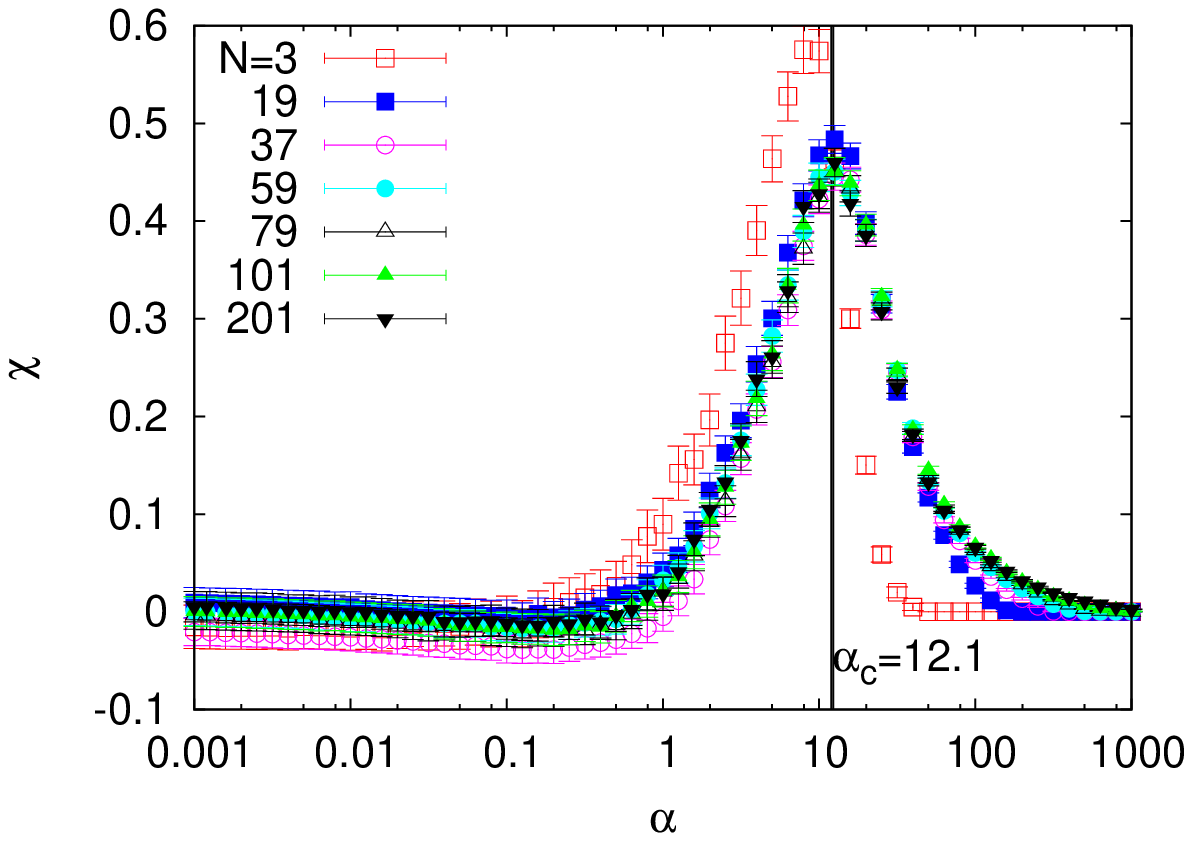}
\end{center}
\caption{Susceptibility $\chi$ as a function of the scaled area
$\alpha=NA$. The cross-over coupling $\alpha_c$ is shown by the vertical
line.}
\eef{susfig}

The reason for this cross-over can be understood from the scatter plot
of $\cas$ versus $\lambda$ measured during the course of the Monte Carlo
run using a value of $\alpha$. Such scatter plots at various $\alpha$ are
shown in \fgn{scatter} for two different $N$. We also show the maximum and
minimum value of Casimir at a fixed $\lambda$, as a function of $\lambda$.
As discussed earlier, the minimum Casimir shows a quasi-periodic behaviour
forming wells with a periodicity $N$.  At large values of $\alpha$,
the representations near the troughs of these wells at small values
of $\lambda$ get populated. The representations within these wells are
sparse, and this discreteness govern the large area behaviour. At very
small area, the most probable $\cas$ moves away from the line of minimum
$\cas$ and remains in a region where one can approximate the distribution
of Casimir by a continuum. The cross-over between the two behaviours is
what shows up as a peak in $\chi$. As discussed in \scn{casimir}, the
Casimir near the troughs at small $\lambda$ is of $\mathscr{O}(N)$, while
the Casimir at very large $\lambda$ is of $\mathscr{O}(N^2)$. As shown
by the dotted line in \fgn{scatter}, this cross-over at $\alpha\approx
12.1$ roughly occurs when the dominant behaviour $\cas$  changes from
$\mathscr{O}(N)$ to $\mathscr{O}(N^2)$.
\bef
\begin{center}
\includegraphics[scale=0.7]{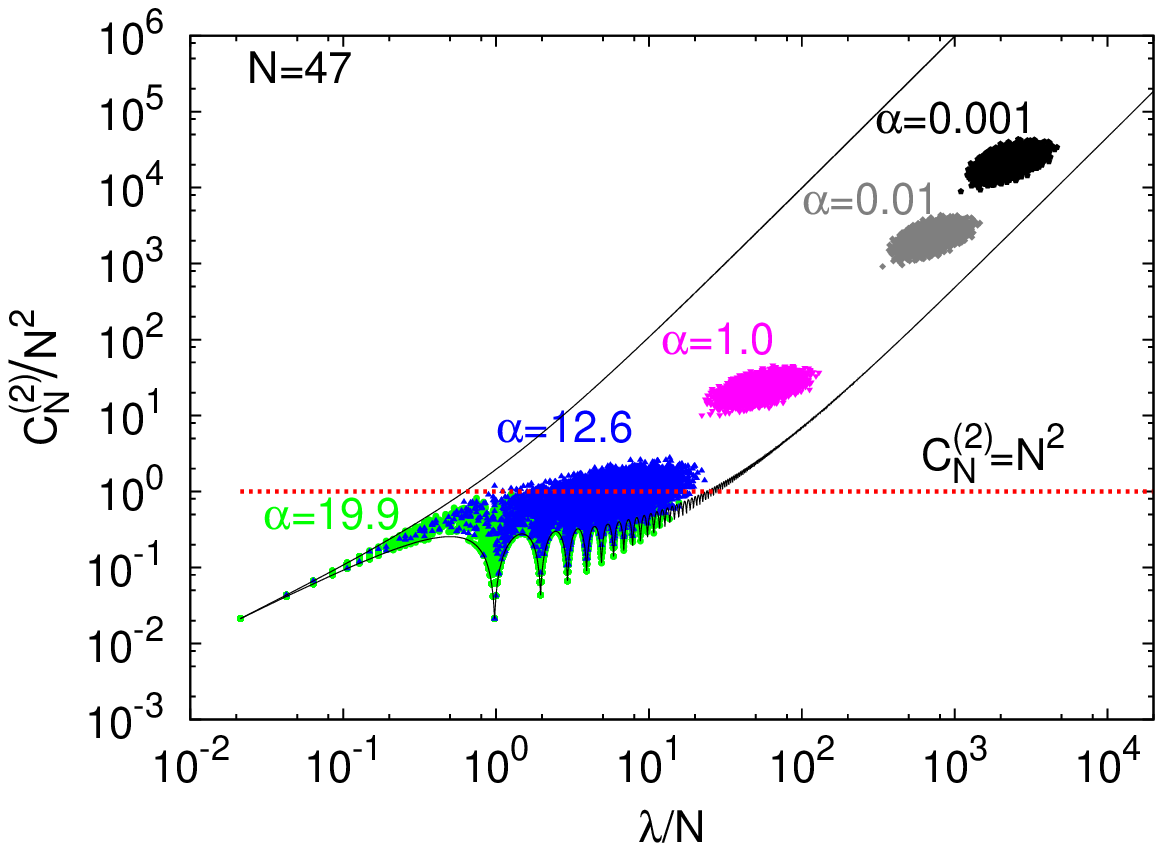}
\includegraphics[scale=0.7]{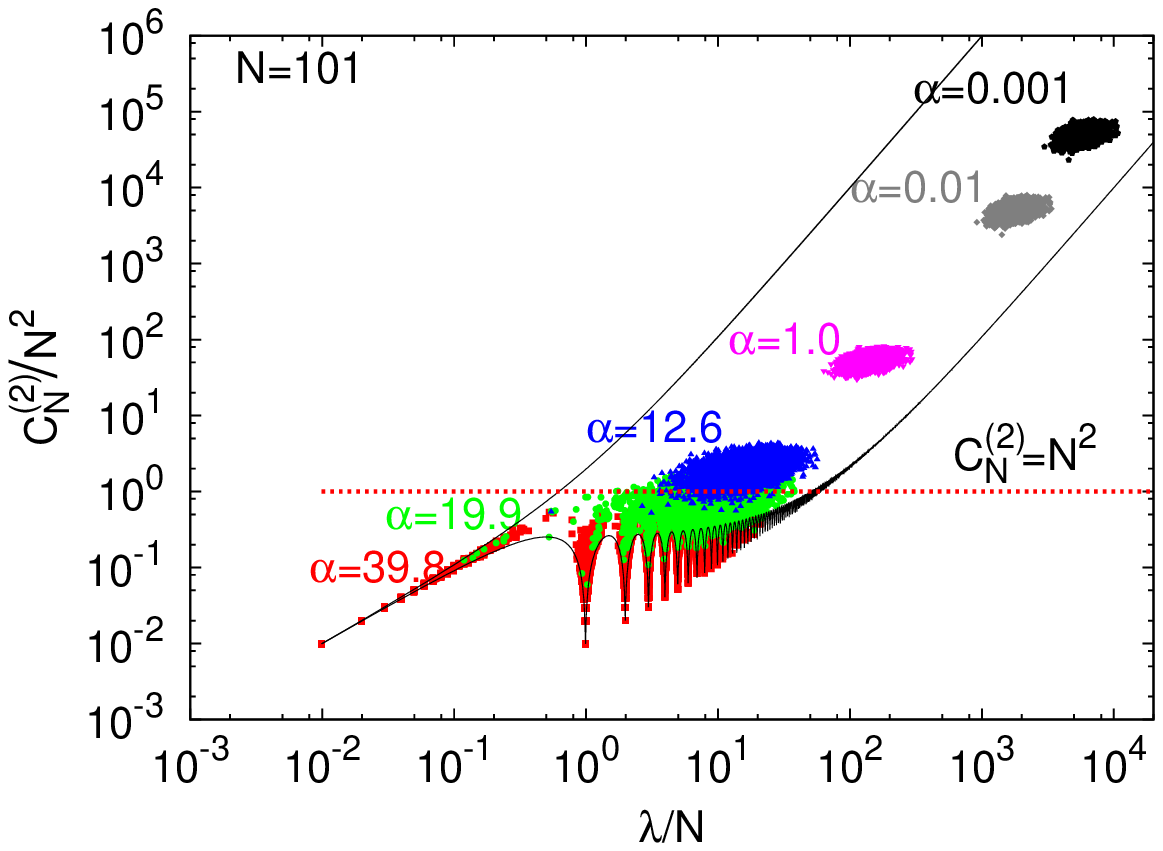}
\end{center}
\caption{Scatter plot of $\cas/N^2$ versus $\lambda/N$ at various area
$A$. The top panel is for $N=47$ and the bottom one for $N=101$. Each
point corresponds to a $\cas$ and $\lambda$ measured in the course
of Monte Carlo simulation at a particular $\alpha$ specified by the
color. The upper and lower solid curves are the maximum and the
minimum value of $\cas$ at a given $\lambda$ respectively.}
\eef{scatter}

\section{Conclusions}

\bef
\begin{center}
\includegraphics[scale=0.8]{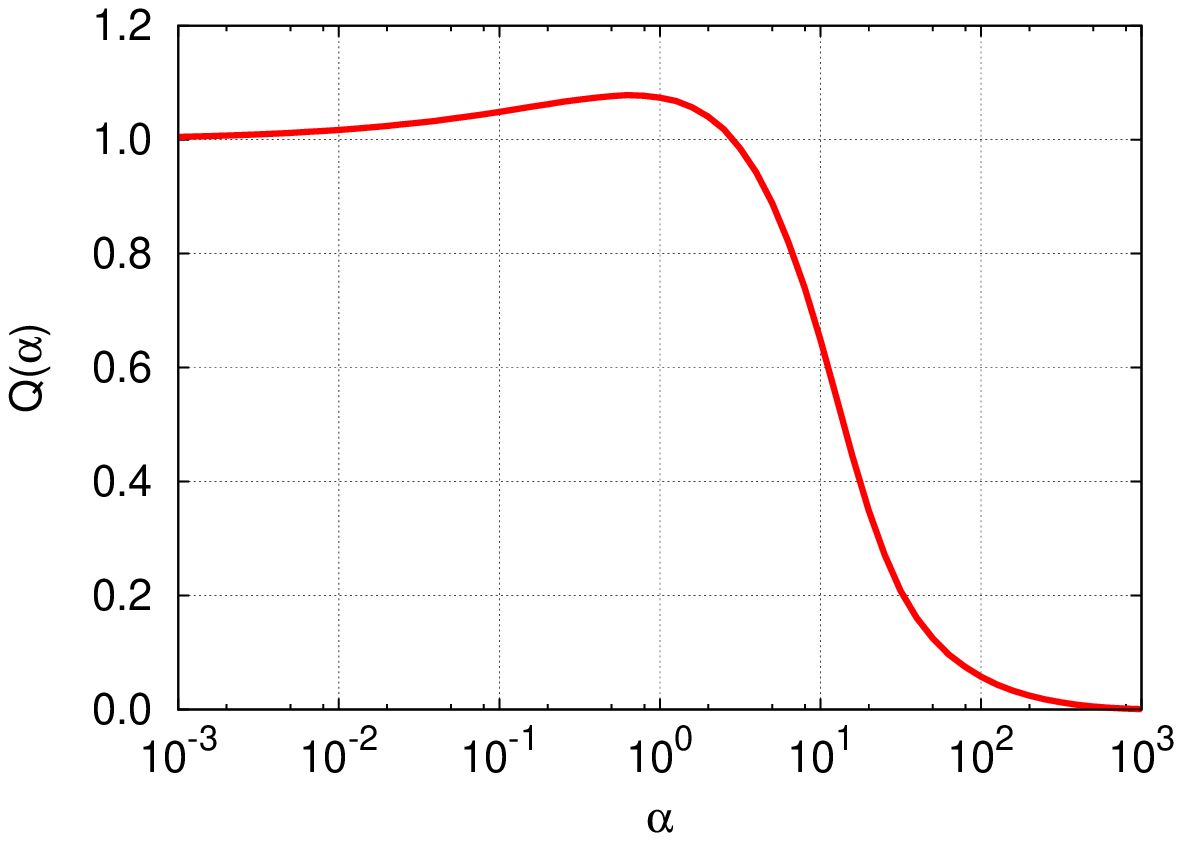}
\end{center}
\caption{The large-$N$ limit of $Q$ as a function of $\alpha$.}
\eef{unicurve}

Yang-Mills theory in two dimensions is always in the confined phase.
We focused on the quantity, $Q=-\frac{2Pl}{(N-1)t}$, to study the equation
of state.  We showed that the equation of state shows a cross-over from strong
coupling (large spatial extent) to weak coupling (small spatial extent)
within the confined phase. Viewed as a function of $\alpha=\frac{lN}{t}$,
$Q(\alpha)$ approaches a universal curve as $N\to\infty$ as shown
in \fgn{unicurve}.  This behavior is similar to the Durhuus-Olesen
transition~\cite{Durhuus:1980nb,Narayanan:2007dv} with the double
scaling limit for the equation of state being $N\to\infty$ and $l\to 0$
(or $t\to\infty$) keeping $\alpha=\frac{lN}{t}$ fixed.  There is a line
of cross-over, $\frac{lN}{t}=\alpha_c$, extending from the origin in
the $\frac{l}{t}$ -- $\frac{1}{N}$  diagram as shown in  \fgn{phase}.
 Well above this line, $Q\ll1$
and it behaves as $\exp(-\sigma A)/N$.  Well below this line, $Q$
is approximately $1$. Depending on the slope, $\alpha$, of the line
along which the $N\rightarrow\infty$ and $\frac{l}{t}\rightarrow0$
limit is taken, the limiting value of $Q$ differs.  Specifically, if
$N\rightarrow\infty$ limit is taken after the $A\rightarrow 0$ limit
is taken, then $Q$ is 1.  When the two limits are reversed, $Q$ becomes
0. Therefore, the cross-over along $AN=\alpha_c$ becomes a first order
transition at vanishing area in the large-$N$ limit.

\bef
\begin{center}
\includegraphics[scale=0.7]{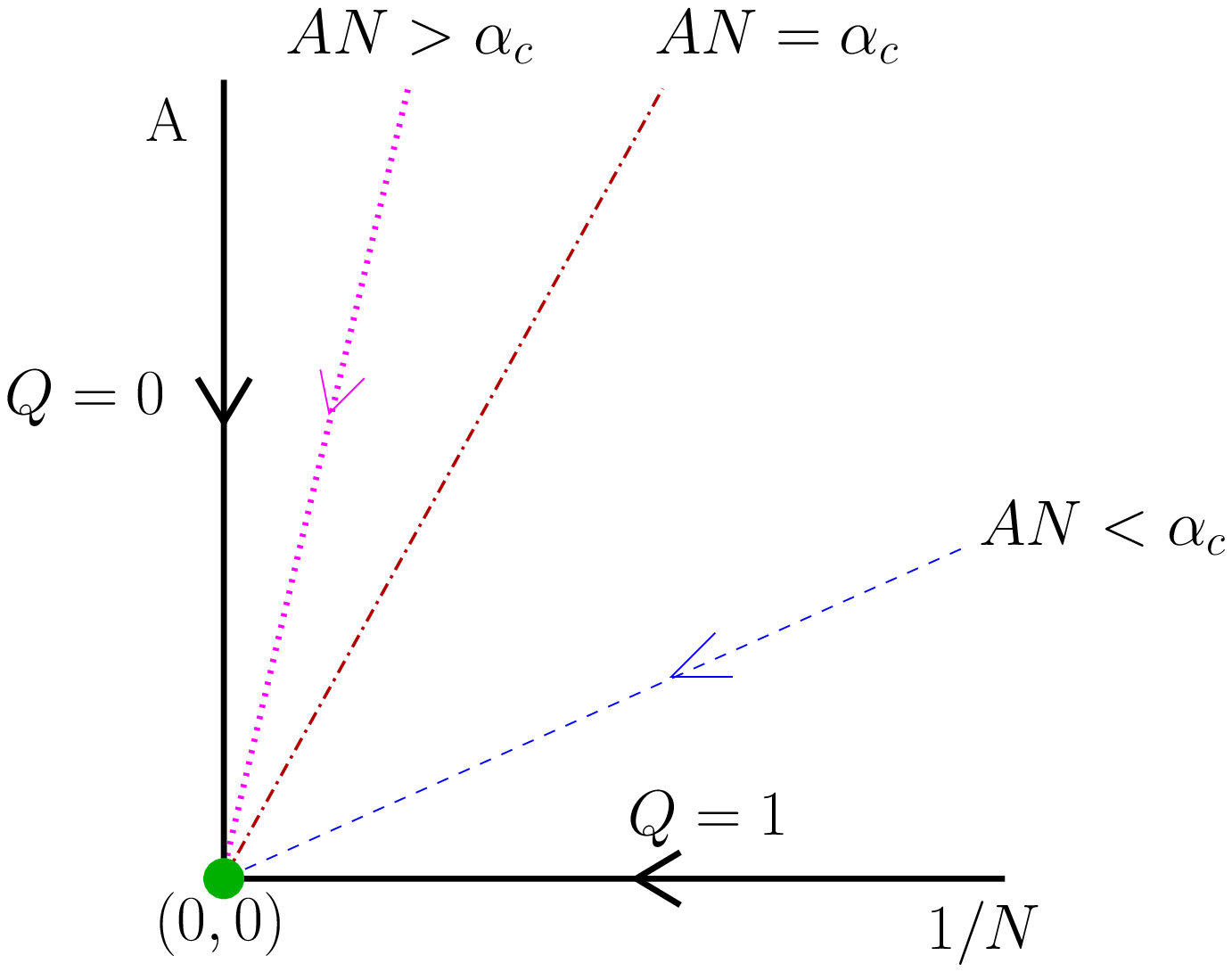}
\end{center}
\caption{Phase diagram. Various approaches to vanishing area at large-$N$
are indicated by lines with arrows. The critical value of the slope
$AN=\alpha_c$ is shown as the dot-dashed line. For values of
$AN\gg\alpha_c$ (the dotted line), $Q$ decays exponentially  with
area. For values of $AN\ll\alpha_c$ (the dashed line), $Q\approx 1$. In
particular, when $A$ is reduced to $0$ after taking the large-$N$ limit
(\ie, along $y$-axis), $Q$ vanishes. When the two limits are interchanged
(\ie, along $x$-axis), $Q$ becomes $1$.}
\eef{phase}

The equation of state in four dimensional Yang-Mills theories for several
different values of $N$ has been recently studied~\cite{Datta:2010sq}. The
pressure is found to be close to zero in the confined phase. In light
of this paper, it would be interesting to perform a careful study of
the equation of state in the confined phase in three and four dimensions
and see if one can see a
cross-over similar to the one seen here in two dimensions.

\acknowledgments

The authors acknowledge partial support by the NSF under grant number
PHY-1205396.

\bibliographystyle{apsrev}
\bibliography{biblio}
\end{document}